\documentclass[aps,prl,twocolumn,superscriptaddress,floatfix,longbibliography]{revtex4-2}
\usepackage{amsmath,amssymb,amsfonts,braket,graphicx,xcolor,mathtools,bm,xfrac,footmisc,paralist,mwe,siunitx}
\usepackage[colorlinks=true,allcolors=blue]{hyperref}
\usepackage[version=4]{mhchem}
\usepackage[T1]{fontenc} % encoding LY1 does not support \dj
\usepackage[varg]{txfonts}
\usepackage{multirow}

\def\equationautorefname~#1\null{Eq.~(#1)\null}
\allowdisplaybreaks

\begin{document}
	\title{Giant Nonvolatile Multistate Resistance with Full Magnetism Control in van der Waals Multiferroic Tunnel Junctions}
	\author{Zhi Yan}
 \email{yanzhi@sxnu.edu.cn}
\affiliation{School of Chemistry and Materials Science $\&$ Key Laboratory of Magnetic Molecules and Magnetic Information Materials of Ministry of Education, Shanxi Normal University, Taiyuan 030031, China}
\affiliation{Research Institute of Materials Science $\&$ Shanxi Key Laboratory of Advanced Magnetic Materials and Devices, Shanxi Normal University, Taiyuan 030031, China}
	\author{Xujin Zhang}
\affiliation{School of Chemistry and Materials Science $\&$ Key Laboratory of Magnetic Molecules and Magnetic Information Materials of Ministry of Education, Shanxi Normal University, Taiyuan 030031, China}
  	\author{Jianhua Xiao}
\affiliation{School of Chemistry and Materials Science $\&$ Key Laboratory of Magnetic Molecules and Magnetic Information Materials of Ministry of Education, Shanxi Normal University, Taiyuan 030031, China}
	\author{Cheng Fang}
\affiliation{School of Chemistry and Materials Science $\&$ Key Laboratory of Magnetic Molecules and Magnetic Information Materials of Ministry of Education, Shanxi Normal University, Taiyuan 030031, China}
\author{Xiaohong Xu}
 \email{xuxh@sxnu.edu.cn}
\affiliation{School of Chemistry and Materials Science $\&$ Key Laboratory of Magnetic Molecules and Magnetic Information Materials of Ministry of Education, Shanxi Normal University, Taiyuan 030031, China}
\affiliation{Research Institute of Materials Science $\&$ Shanxi Key Laboratory of Advanced Magnetic Materials and Devices, Shanxi Normal University, Taiyuan 030031, China}
	\date{\today}
	
	\begin{abstract}
Ferroelectric polarization switching in electrically controlled van der Waals multiferroic tunnel junctions (vdW-MFTJs) causes atomic migration, compromising device stability and fatigue resistance. Here we propose a fully magnetically controlled vdW-MFTJ based on a \(\mathrm{CrBr_3/MnPSe_3/CrBr_3}\) vertical heterostructure, which achieves ferroelectric polarization reversal without relying on atomic migration driven by inversion symmetry breaking. Using first-principles calculations, we investigate the spin-polarized quantum transport properties of the proposed structure. By integrating asymmetric PtTe$_2$/alkali-metal (Li/Na/K)-doped/intercalated CrBr$_3$ electrodes, the device demonstrates exceptional performance, with a maximum tunneling magnetoresistance (TMR) exceeding $8.1\times10^5$\% and tunneling electroresistance (TER) reaching 2499\%, while the spin-filtering channels can be flexibly controlled by the magnetization direction of the magnetic free layer, achieving perfect spin-filtering over a broad bias voltage range. Applying an external bias voltage further enhances these metrics, increasing TMR to $3.6\times 10^7$\% and TER to 9990\%. Notably, a pronounced negative differential resistance (NDR) effect is observed, yielding an unprecedented peak-to-valley ratio (PVR) of $9.55\times10^9$\%, representing the highest value reported for vertical tunnel junctions. 
These extraordinary characteristics highlight the potential of vdW-MFTJs for ultra-efficient electronic switching, a key feature for next-generation spintronic devices. Our findings provide a solid theoretical foundation for designing and developing high-performance magnetic storage and logic technologies.

	\end{abstract}
	
 	\maketitle
	
\textit{Introduction}.---The advancement of memory and logic technologies increasingly hinges on multifunctional devices capable of integrating diverse physical properties~\cite{vzutic2004spintronics,wolf2001spintronics}. Among these, van der Waals (vdW) heterostructures—assembled from atomically thin layers with weak interlayer coupling—have emerged as prime candidates for next-generation electronics due to their unparalleled tunability and intrinsic low-dimensional nature~\cite{geim2013van,liu2016van,novoselov20162d,zhai2021electrically,yan2020significant,yan2021barrier}. In particular, vdW multiferroic tunnel junctions (vdW-MFTJs) have garnered attention for data storage applications, owing to their ability to exhibit multiple nonvolatile resistance states driven by the coexistence of ferroelectricity and ferromagnetism. Their adjustable thickness and interfacial properties offer a versatile platform for engineering quantum transport phenomena~\cite{lin2022engineering,duong2017van}. Despite extensive studies~\cite{dong2023voltage,yan2024giantelectrode,yan2022giant,feng2024van,su2020van,xie2023giant,zhu2023large,zhang2024spin,zhang2023multilevel}, most existing vdW-MFTJs treat ferroelectricity and ferromagnetism as independent effects, mechanically combining them without achieving direct magnetoelectric coupling. This limitation results in suboptimal device performance and unnecessary energy dissipation, highlighting the need for innovative designs that achieve true magnetoelectric coupling.

Magnetoelectric coupling in vdW-MFTJs can be categorized into two mechanisms: (i) ferroelectric-controlled magnetization direction, which is completely dependent on the electric field, and (ii) magnetization-controlled ferroelectric polarization reversal, which relies entirely on the magnetic field. To date, only two studies have reported fully electrically controlled vdW-MFTJs~\cite{yang2023full,yu2023fully}, where the reversal of ferroelectric polarization is accompanied by atomic migration, compromising the device's fatigue resistance and stability. In contrast, magnetically controlled ferroelectric polarization reversal remains largely unexplored, representing a significant gap in research.

This gap highlights a key challenge in achieving genuine magnetoelectric coupling-overcoming the intrinsic decoupling between spin and charge degrees of freedom~\cite{spaldin2019advances,tokura2014multiferroics,fiebig2016evolution}. 
Recent advances in two-dimensional (2D) multiferroic systems have explored two primary classes of coupling mechanisms~\cite{zhong2020room,liu2020magnetoelectric,zhao2024realization}: type-I multiferroics, such as FeI\(_2\)/In\(_2\)Se\(_3\)~\cite{sun2019valence}, CrI\(_3\)/Sc\(_2\)Co\(_2\)~\cite{zhao2019nonvolatile}, and charged CrBr\(_3\)~\cite{huang2018prediction}, and type-II multiferroics, including Hf\(_2\)VC\(_2\)F\(_2\)~\cite{zhang2018type}, VOI\(_2\)~\cite{ding2020ferroelectricity}, and MnCl\(_2\)~\cite{prayitno2019first}. However, these systems face critical limitations: weak magnetoelectric coupling and the confinement of ferroelectric polarization to in-plane directions, which is less practical for device integration. 
Recently, a CrBr$_3$/MnPSe$_3$/CrBr$_3$ heterostructure was proposed~\cite{li2023spin}, demonstrating spin-induced out-of-plane ferroelectricity and robust magnetoelectric coupling. This heterostructure shows promise as a building block for vdW-MFTJs. However, achieving enhanced spin filtering, tunable resistance ratios, and nontrivial quantum effects in vdW-MFTJs remains a formidable challenge.

    \begin{figure*}[htb!]
	\centering
	\includegraphics[width=18cm,angle=0]{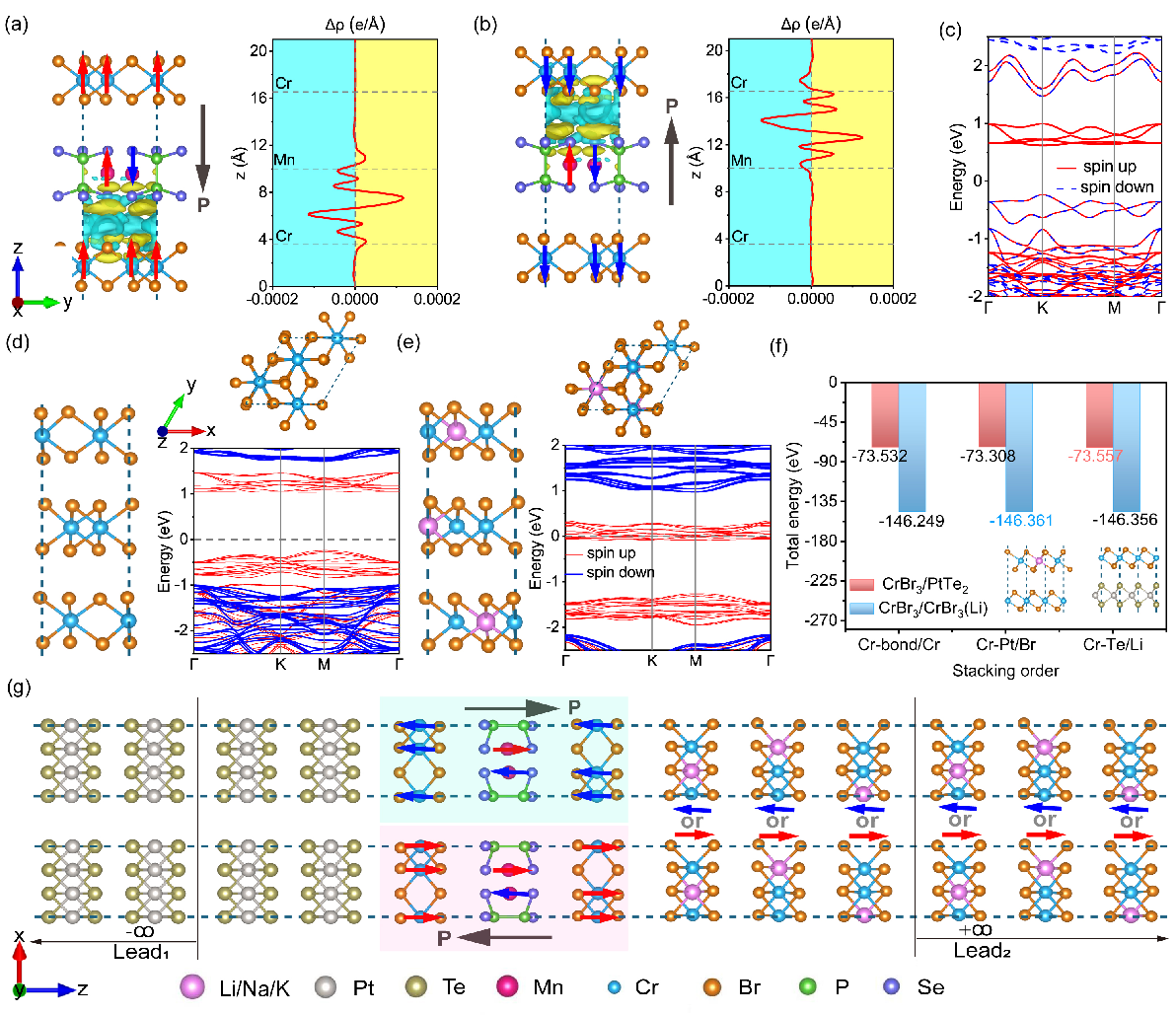}
	\caption{(a) and (b) Side views of the CrBr$_3$/MnPSe$_3$/CrBr$_3$ heterostructure showing the differential charge density between the out-of-plane ferroelectric and non-polarized states, with plane-averaged plots. Electron accumulation and depletion are indicated by yellow and cyan isosurfaces, set at $1.65 \times 10^{-5}$~$e$/Bohr$^3$. (c) Electronic band structure of the CrBr$_3$/MnPSe$_3$/CrBr$_3$ heterostructure. 
        (d) and (e) Side and top views of the bulk CrBr$_3$ crystal, the Li-intercalated CrBr$_3$ structure, and their corresponding band structures. 
        (f) Total energy at the CrBr$_3$/PtTe$_2$ and CrBr$_3$/BrCr$_3$(Li) interfaces for various stacking sequences, with insets showing the optimal stacking configurations. (g) Schematic diagram of the PtTe$_2$/CrBr$_3$/MnPSe$_3$/CrBr$_3$/CrBr$_3$(Li/Na/K) vdW-MFTJs device model, periodic in the $xy$-plane, with current flowing along the $z$-direction.
    }
	\label{Fig1}
\end{figure*}

In this Letter, we present a fully magnetically controlled vdW-MFTJ design based on a CrBr$_3$/MnPSe$_3$/CrBr$_3$ vertical heterostructure. This novel design enables ferroelectric polarization reversal without atomic migration, circumventing the problems associated with inversion symmetry breaking found in traditional electrically controlled systems.  
Employing first-principles calculations, we investigated its spin-polarized quantum transport properties and demonstrated remarkable performance enhancements through the integration of asymmetric PtTe$_2$/alkali-metal (Li/Na/K)-doped/intercalated CrBr$_3$ electrodes. The proposed design exhibits exceptional tunneling magnetoresistance (TMR) and tunneling electroresistance (TER), reaching 8.1$\times$10$^5$\% and 2499\%, respectively, along with perfect spin-filtering effects over a broad bias voltage range. Furthermore, applying an external bias voltage amplifies these metrics to unprecedented levels, with TMR and TER increasing to 3.6$\times$10$^7$\% and 9990\%, respectively. 
Notably, the device displays a pronounced negative differential resistance (NDR) effect, characterized by a peak-to-valley ratio (PVR) as high as 9.55$\times$10$^9$\%, the largest reported in vertical tunnel junctions. These extraordinary properties underscore the transformative potential of vdW-MFTJs for ultra-efficient electronic switching, laying a robust theoretical foundation for next-generation spintronic devices and advancing high-performance magnetic storage and logic technologies.

    \begin{table*}[htp!]
	\centering
 	\renewcommand{\arraystretch}{1.4}
	\caption{Calculated the spin-resolved electron transmission $T_{\uparrow}$ and $T_{\downarrow}$, TMR, TER, and SIE of PtTe$_2$/CrBr$_3$/MnPSe$_3$/CrBr$_3$/CrBr$_3$(Li/Na/K) vdW-MFTJs at equilibrium state.}
	\label{table1}
 \resizebox{\linewidth}{!}{
\begin{tabular}{c c c c c c c c c c c c c}
\hline\hline
\multicolumn{2}{c}{\raisebox{-0.9em}{Dopant Atom}} & \multicolumn{2}{c}{\raisebox{-0.9em}{Polarization}} & \multicolumn{4}{c}{PC state (M${\uparrow\uparrow}$)} & \multicolumn{4}{c}{APC state (M${\uparrow\downarrow}$)} & \multirow{2}{*}{\large$\substack{\text{TMR}\\}$} \\ 
\cline{5-12}
\multicolumn{2}{c}{\raisebox{0.5em}{\ }} & \multicolumn{2}{c}{\raisebox{0.5em}{and Ratio}} & $T_{\uparrow}$ & $T_{\downarrow}$ & $T_\text{tot}=T_{\uparrow}+T_{\downarrow}$ & SIE & $T_{\uparrow}$ & $T_{\downarrow}$ & $T_\text{tot}=T_{\uparrow}+T_{\downarrow}$ & SIE & \\
\hline
\multicolumn{2}{c}{Li} & \multicolumn{2}{c}{$\mathrm{P} \rightarrow$} & {$9.75\times10^{-7}$} & {$3.22 \times 10^{-24}$} & {$9.75 \times 10^{-7}$} & 100 \% & {$8.45 \times 10^{-24}$} & {$7.33 \times 10^{-10}$} & {$7.33 \times 10^{-10}$} & 100 \% & $132915 \%$ \\
\cline{3-13}
\multicolumn{2}{c}{\ } & \multicolumn{2}{c}{$\mathrm{P} \leftarrow$} & {$3.29 \times 10^{-24}$} & {$2.29 \times 10^{-7}$} & {$2.29 \times 10^{-7}$} & 100 \% & {$2.82 \times 10^{-11}$} & {$7.86 \times 10^{-24}$} & {$2.82 \times 10^{-11}$} & 100 \% & $811957 \%$ \\
\cline{3-13}
\multicolumn{2}{c}{\ } & \multicolumn{2}{c}{TER}& \multicolumn{4}{c}{326\%} & \multicolumn{4}{c}{2499\%} &  \\
\hline
\multicolumn{2}{c}{Na} & \multicolumn{2}{c}{$\mathrm{P} \rightarrow$} & {$3.26 \times 10^{-10}$} & {$4.40 \times 10^{-24}$} & {$3.26 \times 10^{-10}$} & 100 \% & {$8.25 \times 10^{-24}$} & {$2.33 \times 10^{-13}$} & {$2.33 \times 10^{-13}$} & 100 \% & $139814 \%$ \\
\cline{3-13}
\multicolumn{2}{c}{\ } & \multicolumn{2}{c}{$\mathrm{P} \leftarrow$} & {$4.49 \times 10^{-24}$} & {$1.71 \times 10^{-10}$} & {$1.71 \times 10^{-10}$} & 100 \% & {$8.49 \times 10^{-14}$} & {$8.91 \times 10^{-24}$} & {$8.49 \times 10^{-14}$} & 100 \% & $201313 \%$ \\
\cline{3-13}
\multicolumn{2}{c}{\ } & \multicolumn{2}{c}{TER}& \multicolumn{4}{c}{91\%} & \multicolumn{4}{c}{174\%} &  \\
\hline
\multicolumn{2}{c}{K} & \multicolumn{2}{c}{$\mathrm{P} \rightarrow$} & {$1.58 \times 10^{-10}$} & {$4.10 \times 10^{-24}$} & {$1.58 \times 10^{-10}$} & 100 \% & {$6.15 \times 10^{-24}$} & {$1.61 \times 10^{-13}$} & {$1.61 \times 10^{-13}$} & 100 \% & $98037 \%$ \\
\cline{3-13}
\multicolumn{2}{c}{\ } & \multicolumn{2}{c}{$\mathrm{P} \leftarrow$} & {$4.20 \times 10^{-24}$} & {$1.52 \times 10^{-10}$} & {$1.52 \times 10^{-10}$} & 100 \% & {$4.61 \times 10^{-14}$} & {$8.54 \times 10^{-23}$} & {$4.61 \times 10^{-14}$} & 100 \% & $329618 \%$ \\
\cline{3-13}
\multicolumn{2}{c}{\ } & \multicolumn{2}{c}{TER}& \multicolumn{4}{c}{3.9\%} & \multicolumn{4}{c}{249\%} &  \\
\hline\hline
\end{tabular}
}
\end{table*}

 	\textit{The establishment of vdW-MFTJs models}.---The ferroelectricity observed in vdW-MFTJs is primarily driven by the intrinsic ferroelectric properties of the materials. Switching the ferroelectric polarization direction results in structural changes, such as sliding~\cite{dong2023voltage} or atomic migration~\cite{yan2022giant,ding2017prediction}, which breaks the inversion symmetry. This process typically requires mechanical energy or an external electric field, making it unsuitable for low-power devices. A recent study~\cite{li2023spin} proposed that spin-driven ferroelectricity, which preserves structural inversion symmetry, can be observed in the vertically stacked CrBr$_3$/MnPSe$_3$/CrBr$_3$ heterostructure. This structure, therefore, holds promise as a core component for vdW-MFTJs, enabling low-power, all-magnetic control of ferroelectric behavior. Figures~\ref{Fig1}(a) and 1(b) illustrate in detail the correlation between the magnetic ordering of the magnetic atoms and the out-of-plane ferroelectricity in this heterostructure. The ferroelectric polarization strength, calculated via the Berry phase method~\cite{king1993theory}, is 0.33 pC/m. When the CrBr$_3$ layers on both sides of MnPSe$_3$ adopt antiferromagnetic ordering, the system exhibits no ferroelectricity, as the geometric and magnetic ordering maintain inversion symmetry. The differential charge density and the planar-averaged plot between the ferroelectric and non-ferroelectric states further confirm the magnetoelectric coupling effect [see Figs.~\ref{Fig1}(a) and 1(b)]. Figure.~\ref{Fig1}(c) shows that the CrBr$_3$/MnPSe$_3$/CrBr$_3$ heterojunction exhibits magnetic semiconductor behavior. The spin-down bands display half-metallic characteristics in the energy range of approximately 0.5 eV to 1.0 eV, making this structure an ideal candidate for use as both the barrier and magnetic layers in vdW-MFTJs. 
    
    The choice of electrodes is crucial for the device performance. In our previous work~\cite{wang2023topological}, we proposed alkali metal (Li/Na/K) doping as a strategy to induce a semiconductor-to-half-metal transition in CrCl$_3$. As shown in Figs.~\ref{Fig1}(d) and 1(e), Li doping shifts the Fermi level of bulk CrBr$_3$ into the spin-down band region, achieving half-metallicity. To ensure experimental feasibility, we also employed a Li intercalation approach, utilizing the well-established ionic intercalation technique~\cite{li2019intercalation}to introduce Li into the vdW gaps of multilayer CrBr$_3$, which similarly results in half-metallic behavior [Fig. S1(b) in the Supplemental Material~\cite{SM}]. 
    Thus, alkali metal (Li/Na/K)-doped or intercalated CrBr$_3$ can serve as an electrode. 
    For the asymmetric electrodes required in ferroelectric tunnel junctions, we chose vdW PtTe$_2$ with metallic properties [see the band structure in Fig.S1(a) in the Supplemental Material~\cite{SM}] as another electrode. 
    The optimized in-plane lattice constant for PtTe$_2$ (3.89 \AA) and CrBr$_3$ (6.42 \AA) correspond to a minimal supercell lattice mismatch of 4.7\% for a  $\sqrt{3}\times\sqrt{3}$@$1\times1$ configuration. 
    We also calculated the stacking configurations for the CrBr$_3$/PtTe$_2$ and CrBr$_3$/CrBr$_3$(Li) heterojunction interfaces [Fig.~\ref{Fig1}(f)], finding that the energetically favorable stackings are Cr-Te and Cr-Br, respectively.
    Based on these results, we constructed a fully vdW heterostructure device, PtTe$_2$/CrBr$_3$/MnPSe$_3$/CrBr$_3$/CrBr$_3$(Li/Na/K), as vdW-MFTJs [see Fig~\ref{Fig1}(g) and Fig.S1(c) in the Supplemental Material~\cite{SM}].

    	\textit{Electronic and transport properties at equilibrium}.---The mechanism of traditional MFTJs relies on external magnetic or electric fields to switch the polarization direction of the magnetic free layer or electrostatic barrier, resulting in four non-volatile resistance states: the magnetic ordering-dependent parallel (PC) and antiparallel (APC) configurations, and the ferroelectric polarization-dependent P$\rightarrow$ and P$\leftarrow$ states. 
        In contrast, due to the unique magnetoelectric coupling in the CrBr$_3$/MnPSe$_3$/CrBr$_3$ heterostructure, the control of these four resistance states can be achieved solely by an external magnetic field, enabling low-power operation of our MFTJs. 
        We first investigated the electronic transport properties of PtTe$_2$/CrBr$_3$/MnPSe$_3$/CrBr$_3$/CrBr$_3$(Li/Na/K) vdW-MFTJs at equilibrium. As summarized in Table \ref{table1}, Li/Na/K doping induces significant TMR and TER effects at equilibrium, with maximum values of 8.1$\times$10$^5$\% and 2499\%, respectively. 
        Furthermore, all four non-volatile resistance states exhibit perfect spin filtering effects, and the direction of the magnetic free layer can flexibly switch the spin channels. 
        The experimentally feasible Li intercalation strategy for CrBr$_3$ is detailed in Table SI of the Supplemental Material~\cite{SM}. 
        Despite a one-order-of-magnitude reduction in TMR/TER, Li intercalation preserves the perfect spin filtering effect, with the TMR/TER still remaining substantial. Therefore, these vdW-MFTJs exhibit giant TMR/TER and perfect spin filtering effects with tunable spin channels, showcasing their significant potential for spintronic applications.

\begin{figure}[htb!]
	\centering
	\includegraphics[width=8.6cm,angle=0]{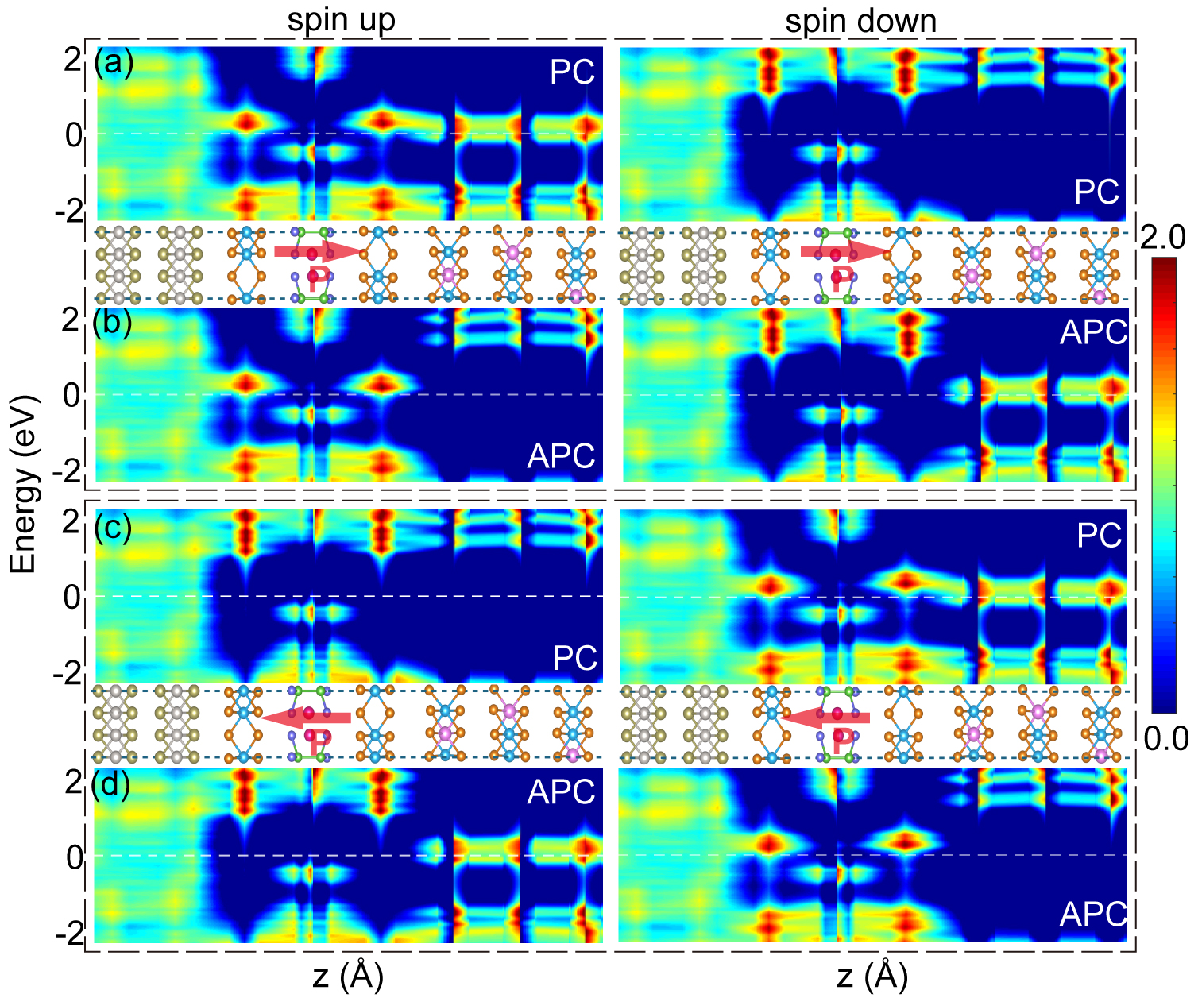}
	\caption{Spin-resolved projected density of states (PDOS) and the corresponding crystal structure of the central scattering region along the transport direction ($z$-axis) of the equilibrium state for PtTe$_2$/CrBr$_3$/MnPSe$_3$/CrBr$_3$/CrBr$_3$(Li) MFTJ. 
        Panels (a) and (b) correspond to the polarization state P$\to$, while panels (c) and (d) correspond to the polarization state P$\gets$. The Fermi level is indicated by a white dashed line.}
	\label{Fig2}
\end{figure}

        The electronic properties of the central scattering region in MFTJs provide insight into the physical mechanisms underlying the non-volatile resistive states. We analyze the Li-doped MFTJs as representatives. 
        As depicted in Fig.~\ref{Fig2}, regardless of the magnetic and ferroelectric configurations, the Fermi level at the MnPSe$_3$ position lies within a non-electronic state region, indicating that electron transport occurs via tunneling. Figure~\ref{Fig2} also demonstrates the typical TMR effect.
        Taking the P$\to$ state as an example, the spin-up density of states (DOS) in Fig.~\ref{Fig2}(a) shows that the CrBr\(_3\)/MnPSe\(_3\)/CrBr\(_3\) heterostructure as a whole and the Li-doped CrBr\(_3\) magnetic free layer can be regarded as majority states due to the significant DOS accumulation near the Fermi level. 
        In contrast, for the spin-down channel, the absence of DOS at the Fermi level for both cases indicates their minority states. 
        In the APC state [Fig.~\ref{Fig2}(b)], spin-up (spin-down) electrons injected from the left electrode pass through the majority (minority) states of the CrBr$_3$/MnPSe$_3$/CrBr$_3$ structure and exit through the minority (majority) states of the Li-doped CrBr$_3$ magnetic electrode. 
        This opposite alignment of electronic states impedes electron transmission, leading to a high-resistance state. 
        In the PC state [Fig.~\ref{Fig2}(a)], the spin-up and spin-down states correspond to majority-to-majority and minority-to-minority transitions, respectively, dictating the low- and high-resistance transport behaviors and yielding a perfect spin-filtering effect, as summarized in Table \ref{table1}.

\begin{figure}[htb!]
	\centering
	\includegraphics[width=8.5cm,angle=0]{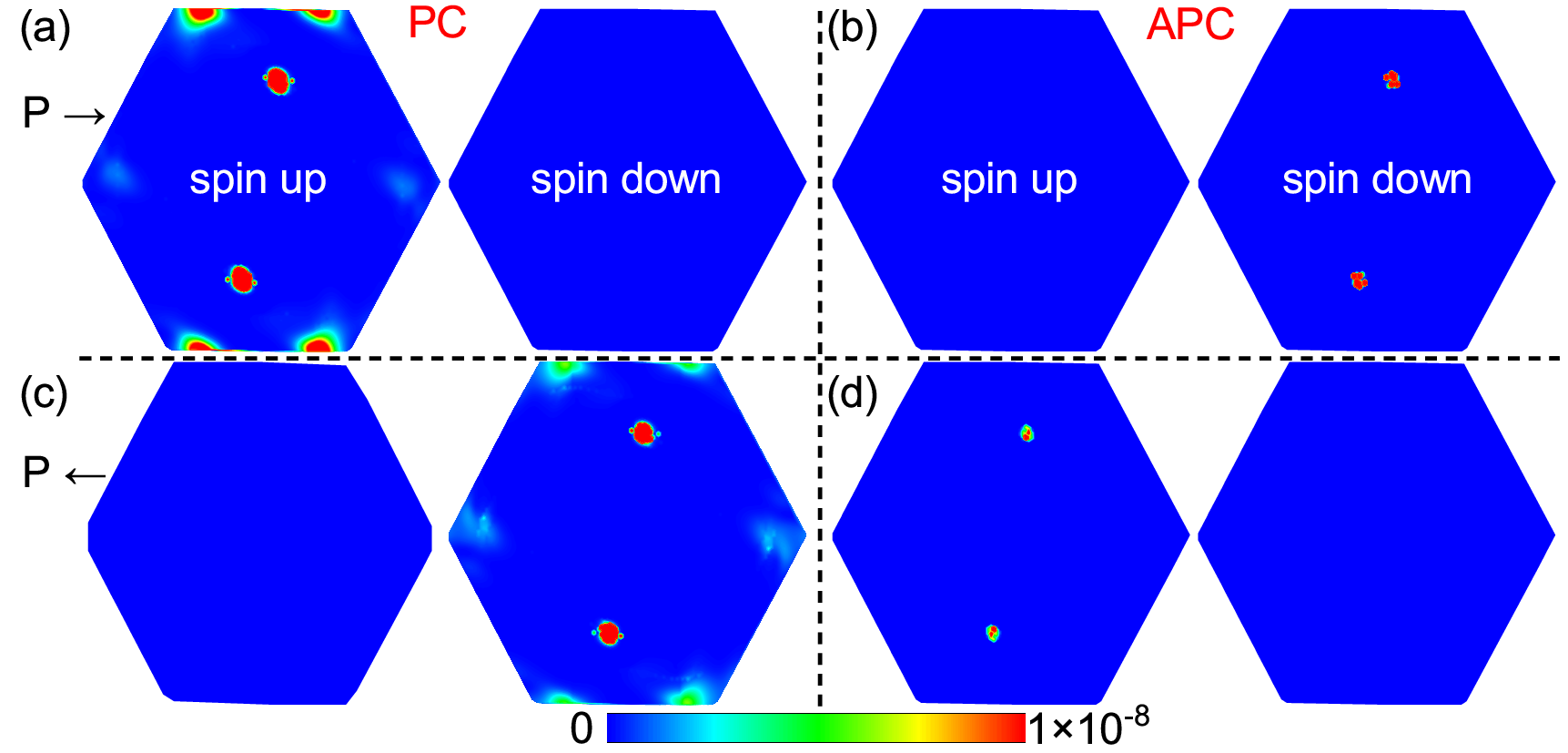}
	\caption{The $k_\Arrowvert$-resolved transmission coefficients of PtTe$_2$/CrBr$_3$/MnPSe$_3$/CrBr$_3$/CrBr$_3$(Li) MFTJs in the 2D Brillouin zone at the Fermi level for the {P$\rightarrow$/P$\leftarrow$} and {PC(M$\uparrow \uparrow$)}/{APC(M$\uparrow \downarrow$)} states.
}
	\label{Fig3}
\end{figure}

    \begin{figure*}[htb!]
	\centering
	\includegraphics[width=18cm,angle=0]{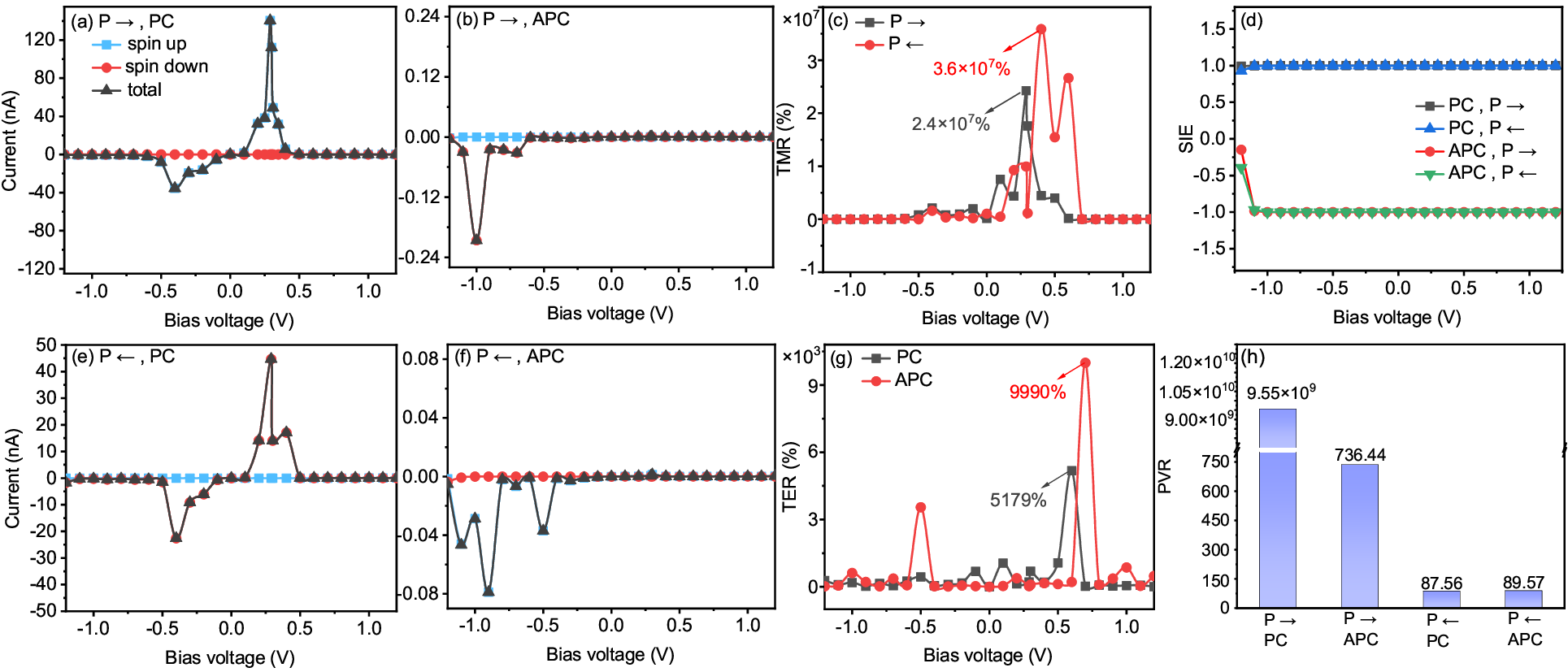}
	\caption{The variation of the spin current [(a), (b), (e), and (f)], TMR ratio (c), TER ratio (g), and spin injection efficiency (SIE) (d) $vs$ the bias voltages and the current peak-to-valley ratio (PVR) (h) in PtTe$_2$/CrBr$_3$/MnPSe$_3$/CrBr$_3$/CrBr$_3$(Li) MFTJs with different polarization directions and magnetic configurations.
}
	\label{Fig4}
\end{figure*}

        To comprehensively analyze the effects of ferroelectric polarization and magnetization alignment on electron transmission, we calculated the $k_\Arrowvert$-resolved transmission coefficients at the Fermi level within the 2D Brillouin zone (2D-BZ). 
        Figures~\ref{Fig3}(a) and 3(b) show that, in the P$\rightarrow$ state, hotspots appear predominantly in the spin-up/spin-down channels of the PC/APC states, with fewer in the spin-down/up channels, indicating a clear spin-filtering effect and magnetization-controlled spin channels. 
        Comparing Figs.~\ref{Fig3}(a) and 3(b), the PC(M$\uparrow \uparrow$) state exhibits significantly more hotspots in the spin-conducting channel than the APC(M$\uparrow \downarrow$) state, suggesting a strong TMR effect. 
        A similar trend is observed in the P$\leftarrow$ state, with reversed spin-filtering channels. 
        A large TER effect is apparent when comparing the hotspot distributions between the P$\rightarrow$ and P$\leftarrow$ states: the distribution in Fig.~\ref{Fig3}(a) is larger than in Fig.~\ref{Fig3}(c), and in Fig.~\ref{Fig3}(b) it is larger than in Fig.~\ref{Fig3}(d). Figures S2 and S3 in the Supplemental Material~\cite{SM} present the PDOS and electronic transmission spectra in the 2D-BZ distribution for Na- and K-doped as well as Li-intercalated electrodes. The analysis above is directly applicable to these cases and will not be repeated here.

        \textit{Transport properties at nonequilibrium}.---In this section, we calculate the bias voltage-dependent (ranging from $-1.2$ to $1.2$ V) spin-polarized current, spin injection efficiency (SIE), TMR, and TER for the MFTJs under four resistance states, as illustrated in Fig.~\ref{Fig4}. From Fig.~\ref{Fig4}(a), one can observe that in the P$\rightarrow$ state with PC magnetic alignment, the total current is primarily contributed by the spin-up current, while the spin-down current is significantly suppressed, demonstrating an intriguing spin-filtering effect. The SIE remains 100\% across the entire bias voltage range [see Fig.~\ref{Fig4}(d)]. 
        Moreover, the total current increases with the bias voltage, then sharply decreases when the bias exceeds $0.29$ V ($-0.4$ V), indicating the presence of a typical negative differential resistance (NDR) effect. 
        The current peak-to-valley ratio (PVR) is crucial for applications of NDR devices. 
        For instance, in high-frequency applications, the PVR determines the efficiency of converting d.c.-to-RF signals~\cite{zhang2023toward}, while in multi-valued logic it defines the voltage range for transitioning to intermediate states~\cite{jo2021recent}. 
        Specifically, the current peaks at $0.29$ V and reaches a minimum at $1.0$ V, with a peak-to-valley difference spanning over ten orders of magnitude, resulting in an exceptionally high PVR of 9.55$\times$10$^9$\% [see Fig.~\ref{Fig4}(h)], the highest value reported to date for vertical tunnel junctions. We have also summarized a series of reported PVR values in Table SII in the Supplemental Material~\cite{SM}. The current highest value, observed in the N-doped monolayer GeS system~\cite{guo2023theoretical}, is two orders of magnitude smaller than ours.
        As displayed in Fig.~\ref{Fig4}(b), in the APC state, the spin-filtering channel is reversed, i.e., spin-up electron transport is blocked, and the NDR effect occurs under negative bias. 
        Comparing Figs.~\ref{Fig4}(a) and 4(b), it is clear that the current in the PC state is significantly larger than in the APC state, indicating a high TMR. 
        For the P$\leftarrow$ state, the behavior is similar to that of the P$\rightarrow$ state. More intriguingly, the distinction lies in the spin-filtering channel being reversed once again as depicted in Figs.~\ref{Fig4}(e) and 4(f), i.e., the PC/APC states conduct spin-down/spin-up currents, respectively. This demonstrates that the spin-filtering channel can be flexibly manipulated among the four resistance states. 
        Furthermore, the current in the P$\rightarrow$ state is substantially larger than in the P$\leftarrow$ state for both magnetic configurations, implying the realization of a large TER. 
        From the I-V curves, the bias-dependent evolution of TMR and TER is derived, as shown in Figs.~\ref{Fig4}(c) and 4(g). The maximum TMR (3.6$\times$10$^7$\%) and TER (9990\%) are achieved at $0.5$ and $0.7$ V, respectively, for the P$\leftarrow$/APC state.

        These nonequilibrium transport properties confirm that the designed vdW-MFTJs not only achieve giant TMR and TER effects but also exhibit perfect spin-filtering with switchable spin channels across a wide bias window in all four non-volatile resistance states. Additionally, they display significant PVR in the NDR effect. These findings highlight the immense potential of our MFTJs in spintronic memory applications.

	\textit{Summary and Discussion}.---In conclusion, the proposed fully magnetically controlled vdW-MFTJ based on the CrBr$_3$/MnPSe$_3$/CrBr$_3$ vertical heterostructure offers a significant advancement in the design of spintronic devices. By enabling ferroelectric polarization reversal without atomic migration, this novel structure overcomes the limitations associated with inversion symmetry breaking in electrically controlled systems. First-principles calculations reveal remarkable enhancements in quantum transport properties, including exceptional TMR and TER, along with perfect spin-filtering effects across a broad voltage range. These performance metrics can be further amplified with external bias, reaching unprecedented levels. Additionally, the pronounced NDR effect, with an exceptional peak-to-valley ratio, highlights the transformative potential of this device for ultra-efficient electronic switching. This work not only provides a solid theoretical foundation for the development of high-performance magnetic storage and logic technologies but also paves the way for the realization of true magnetoelectric coupling in vdW-MFTJs, a crucial step for advancing next-generation spintronic devices. 
    
    The observed magnetoelectric coupling arises from the intrinsic interplay between ferromagnetic CrBr$_3$ and antiferromagnetic MnPSe$_3$, amplified by the unique electronic structure at the heterointerfaces. This coupling enables precise modulation of the tunneling conductance via polarization switching, which is unattainable in conventional vdW-MFTJ designs that primarily rely on independent ferroelectric or magnetic states. Incorporating asymmetric electrodes further enhances the device performance by optimizing the spin polarization and charge transfer at the interfaces. Future work could explore alternative material combinations and stacking configurations to further improve device efficiency and expand operational versatility, paving the way for next-generation spintronic technologies.
	
	\textit{Acknowledgement}.---This work was supported by the National Natural Science Foundation of China (No. 12304148), the National Natural Science Foundation of China Regional Innovation and Development Joint Fund Key Program (No. U24A6002), and the Shanxi Natural Science Basic Research Program (No. 202203021222219). Z.Y. was partially supported by the HZWTECH program (HZWTECH-PROP), and X.Z. by the Graduate Science and Technology Innovation Project Foundation of Shanxi Normal University (No. 2024XSY42).

\bibliography{main}% Produces the bibliography via BibTeX
\end{document}